\documentclass[prb,aps,showpacs,twocolumn,superscriptaddress,10pt,amsmath,amssymb]{revtex4-1}
\usepackage{epsfig}
\usepackage{amsmath}
\usepackage{amssymb}

\newcommand{\epso}{\epsilon_0}
\newcommand{\epsr}{\epsilon_r}
\newcommand{\epsrz}{$\epsilon_r(z)$}
\newcommand{\eq}[1]{~(\ref{eq:#1})}
\newcommand{\eqs}[2]{~(\ref{eq:#1}-\ref{eq:#2})}
\newcommand{\rhoav}[1]{\overline{\delta \rho}(#1)}
\newcommand{\Eav}{\overline{E}}
\newcommand{\Ee}{E_{ext}}
\newcommand{\vhav}[1]{\overline{\delta V_H}(#1)}
\newcommand{\sigmau}{$e\times 10^{12}/\mathrm{cm}^2$}
\newcommand{\sigmauC}{$\mathrm{C}\times 10^{-2}/\mathrm{m}^2$}

\makeatletter 

\begin{document}

\title{{\it Ab initio} study of the relationship between spontaneous polarization and $p$-type doping in quasi-freestanding graphene on H-passivated SiC surfaces}

\author{J. S\l awi\'{n}ska}
\affiliation{Instituto de Ciencia de Materiales de Madrid, ICMM-CSIC, Cantoblanco, 28049 Madrid, Spain.}
\affiliation{Department of Solid State Physics, University of Lodz, Pomorska 149/153, 90236 \L \'{o}d\'{z}, Poland}

\author{H. Aramberri}
\affiliation{Instituto de Ciencia de Materiales de Madrid, ICMM-CSIC, Cantoblanco, 28049 Madrid, Spain.}

\author{M.C. Mu\~noz}
\affiliation{Instituto de Ciencia de Materiales de Madrid, ICMM-CSIC, Cantoblanco, 28049 Madrid, Spain.}

\author{J. I. Cerd\'a}
\affiliation{Instituto de Ciencia de Materiales de Madrid, ICMM-CSIC, Cantoblanco, 28049 Madrid, Spain.}
\email[]{}
\date{\today}
\begin{abstract}
Quasi-free standing graphene (QFG) obtained by the intercalation of a
hydrogen layer between a SiC surface and the graphene is recognized as an
excellent candidate for the development of graphene based technology. In
addition, the recent proposal of a direct equivalence
between the $p$-type doping typically found for these systems and the 
spontaneous polarization (SP) associated to the particular SiC polytype,
opens the possibility of tuning the number of carriers in the Dirac cones
without the need of external gate voltages. However, first principles
calculations which could confirm at the atomic scale the effect of the SP 
are lacking mainly due to the difficulty of combining a bulk property
such as the SP with the surface confined graphene doping.
Here we develop an approach based on standard density functional theory 
(DFT) slab calculations in order to quantify the effect of the SP on the
QFG doping level. First, we present an accurate scheme to estimate
the SPs by exploiting the dependence of the slab's dipole moment with its
thickness. Next, and in order to circumvent the DFT shortcomings associated to 
polar slab geometries, a double gold layer is attached at the C-terminated 
bottom of the slab which introduces a metal induced gap state that pins the
chemical potential inside the gap thus allowing a meaningful comparison
of the QFG dopings among different polytypes. Furthermore, the slab dipole
can be removed after adjusting the Au-Au interlayer distances.
Our results confirm that the SP does indeed induce a substantial
$p$-doping of the Dirac cones which can be as large as a few hundreds of
meV depending on the hexagonality of the polytype. 
The evolution of the doping with the slab thickness reveals
that several tens of SiC bilayers are required to effectively remove the
depolarization field and recover the macroscopic regime whereby the
graphene doping should equal the SP.
\end{abstract}

\maketitle

\section{Introduction}
Obtaining quasi-freestanding graphene (QFG) from epitaxial graphene (EG) on the
SiC(111)/(0001)-face of silicon carbide (SiC) via intercalation of a H 
layer~\cite{cubic, riedl, riedl_review, virojanadara} has recently become a 
promising route to fabricate large area graphene of high-quality. Due to the
reduced graphene-substrate interaction, QFG presents improved carrier mobilites
as compared to graphene on the buffer layer (2000-3000 vs 700-900 cm$^{2}$/Vs)
together with a weak dependence on temperature~\cite{raman,jjap}.
Excellent performance of QFG-based transistors has been already reported, for 
instance, by Robinson {\it et al}~\cite{transistors}. Furthermore, as observed 
in scanning tunneling microscope (STM) experiments~\cite{goler} QFG is hardly 
corrugated and almost defect-free, which makes it an interesting system in
the context of many-body theories in "2+1" dimensions. The electronic structure
of QFG has been intensively studied during the last few years under different 
experimental techniques~\cite{riedl,riedl_review,cubic,virojanadara,forti,plasmons,transistors,jjap,goler,raman,elpho,hibino} 
and a general consensus has been reached regarding the preservation of the 
linear Dirac cones. The most intriguing property, however, is the $p$-type 
doping consistently found in QFG, which is at striking contrast with the large 
$n$-type doping values ($\sim e\times10^{13}$cm$^{-2}$) generally measured for 
EG~\cite{riedl_review, raman}. In Table~\ref{alldope} we present a summary of
the doping charges, $\delta\sigma$, or alternatively the Dirac point (DP) shifts
with respect to the chemical potential, $\Delta$DP, reported in some
representative works. Indeed, an ample range of doping levels has been measured
for similarly prepared QFG samples~\cite{riedl,riedl_review,goler,forti,transistors,raman,jjap,elpho,plasmons,virojanadara}, 
attaining values larger than 300~meV~\cite{newrajput} or even small $n$-type 
doping~\cite{cubic}. Furthermore, the values presented in the table hint
certain correlation between the doping level and the hexagonality of the
underlying SiC polytype; while cubic 3C-SiC(111) samples show relatively small 
$n$-type doping, all 6H- and 4H-SiC(0001) samples are $p$-doped. Very recently
Mammadov~{\it et al}~\cite{newrms} performed a systematic study of this
correlation to find that the graphene doping was around 1.5~times larger in
4H than in 6H samples (see Table~\ref{alldope}), regardless of the substrate's
doping level. Notably, similar doping levels are found for bilayer~\cite{newrms}
and even trilayer~\cite{newtrilayer} graphene, since in such multilayer systems
the spacing between the occupied $\pi$ bands is sufficiently large so that
only the uppermost one becomes doped~\cite{newrms}.

Although the possibility of tuning the doping of the graphene layer across such
a wide energy range (equivalent to hole concentrations of up to 
2~$e\times 10^{13}$ cm$^{-2}$) is a key issue for the fabrication of QFG-based 
elements,\cite{schottky} the origin of the doping has remained controversial
in many recent works~\cite{newrajput,newpasquarello,newth,fortinew,newtrilayer, nitro}. 
The main sources of graphene doping are recognized to be: 
(i) self-doping induced by intrinsic defects~\cite{defects}, 
(ii) the substrate bulk doping \cite{newrms} and, (iii) the spontaneous 
polarization (SP) associated to the particular SiC polytype employed as 
substrate. The former stems from charge accumulation in the vicinity of a
defect (vacancy or adsorbate); its electronic and magnetic properties 
have been characterized in detail from the theoretical side for both 
free-standing graphene (FG)~\cite{yazyev,yazyev2,ourcarbon} and 
QFG~\cite{ourcarbon}. The second arises from the details of the band bending
at the surface and will in general depend on the nature and concentration of
the bulk dopants which determine the location of the chemical potential 
within the gap (such mechanism is thought to be the cause of the mild
$n$-type doping found for QFG on 3C-SiC(111) samples\cite{newrms}). 
Last, the influence
of the SP on the doping was first proposed by Ristein, Mammadov and Seyller 
(RMS) based on macroscopic dielectric theory~\cite{pdope} (a similar analysis 
was later presented in Ref.~[\onlinecite{davydov}]) and next corroborated
experimentally~\cite{newrms} (Table~\ref{alldope}). This so called spontaneous
polarization doping model assumes that the SP creates a {\it pseudo-charge} at 
the surface equivalent to an acceptor layer which should induce considerable 
doping charges of the order of 6-9~\sigmau\ for the most common 6H/4H-SiC(0001)
substrates. Furthermore, 
since the SP along the (0001) direction remains negative for all hexagonal SiC
polytypes, only $p$-doping should be induced at the Si terminated QFG systems, 
in accordance with the experimental observations listed in Table~\ref{alldope}.

Generally, the SP occurs in dielectric crystals with a distribution of dipoles
along the surface normal such as those created at planar stacking defects (SDs);
that is, when a stacking sequence is altered with respect to that in an ideal 
cubic crystal~\cite{qteish,park,pdope}. In the bulk phase periodic boundary 
conditions impose a vanishing net electric field across the unit cell.
However, at surfaces translation symmetry is broken 
and the dipoles may add up generating an uncompensated polarization field.
As a consequence, the electrostatic (Hartree) 
potential, $V_H$, will raise or lower leading to an electrostatically
unstable surface unless a source of hole or electron trapping is attached
to it --the QFG layer, in our case, whose Dirac cones will end up
$p$- or $n$-type doped, respectively.
The SP is an intrinsic property of
the dielectric characterized by its density of dipoles and their sign and 
magnitude, but the final value of the band bending at the surface will be 
also determined by the density of states (DOS) of the compensating charges
at the 2D surface bands --for QFG, ideally
linear in energy-- as well as by the screening capabilities of the free carriers
present in the dielectric which, in turn, depend on the temperature and the 
nature and concentration of impurities (bulk dopants)~\cite{newrms}.

\begin{table}
\caption{Summary of experimental doping values reported for graphene on 
         hydrogenated SiC(111)/(0001) substrates employing different techniques.
         $\Delta$DP expressed in meV and $\delta\sigma$ in 
         $e\times10^{12}$cm$^{-2}$.
         \label{alldope}}
\begin{tabular}{cccccc}
\hline \hline
   SiC  Politype  & $\Delta$DP  & $\delta\sigma$&Technique& Reference \\ \hline
   4H, 6H                &  100        &               &   ARPES & [\onlinecite{riedl,riedl_review}] \\
   4H, 6H\footnote{cubic terminated}& $\sim$150 &  2.3 &   ARPES & [\onlinecite{forti}] \\
   3C                    &  $\sim$-100 &  $\sim$-1.0   &   ARPES & [\onlinecite{cubic}]\\
                         &             &       2.0     &   Hall  & [\onlinecite{transistors}] \\
   6H                    & 13          &               &   STS   & [\onlinecite{goler}] \\
   6H                    &             &  5.0-6.5      &   Hall  & [\onlinecite{raman}] \\
   4H                    &             &  15.0-20.0    &   Hall  & [\onlinecite{nitro}] \\
   4H                    &             &  20.0         &   Hall  & [\onlinecite{newexp}] \\
                         &             &  5.0          &   ARPES & [\onlinecite{elpho}] \\
   6H                    & 320         &               &  STS & [\onlinecite{newrajput}] \\
   6H                    & 320         &  5.0          &   STS   & [\onlinecite{schottky}] \\ 
   6H\footnote{$n$-type doped} & 240         &  4.2          &   ARPES   & [\onlinecite{newrms}] \\ 
   6H\footnote{semi-insulating}& 280         &  6.2          &   ARPES   & [\onlinecite{newrms}] \\ 
   4H$^{b}$ & 300         &  6.9          &   ARPES   & [\onlinecite{newrms}] \\ 
   4H$^{c}$ & 340         &  8.6          &   ARPES   & [\onlinecite{newrms}] \\
   3C$^{b}$ & -100        &  -0.7         &   ARPES   & [\onlinecite{newrms}] \\
\hline \hline
\end{tabular}
\end{table}

Although macroscopic theory predicts a direct relationship between the SP and the 
doping charge in the QFG layer~\cite{pdope}, it is not obvious if it still holds at the
nanoscale. From the theoretical side, this represents a challenging task; despite 
several groups have reported {\it ab initio} SP estimates for the most common 2H-, 4H-
and 6H-SiC(0001) polytypes following different approaches~\cite{qteish,sp2physicab,park,brandino},
no equivalent calculation has been attempted to date addressing its impact on the QFG 
doping. The difficulties associated to the slab geometries typically employed in 
{\it ab initio} studies of surfaces are various. First, one needs to reconcile a surface
property such as the graphene doping with a bulk property such as the SP. Second, the 
polar character of the SiC(0001) slab~\cite{noguera} leads to an
uncompensated {\it compositional} polarization which will affect the calculated
doping levels. Last, a reference chemical potential in the slab 
independent of the selected polytype is required to render any differences in the 
graphene doping meaningful. The aim of this study is precisely to develop a framework
which circumvents these drawbacks allowing a precise determination of the QFG doping 
induced by the SP and, ultimately, establish their relationship. 

In the spirit of Fu~{\it et al}~\cite{fu} (FYRR) and Shi and Ramprasad~\cite{sr}
(SR) our scheme involves two dimensional slabs which can be solved by standard 
approaches such as density functional theory (DFT). 
We first generalize the SR slab formalism in order to estimate the bulk SP of 
the different SiC polytypes including the appropiate corrections required by 
the polar character of the SiC(0001). In doing so, we arrive at a general
and simple, yet self-contained, expression relating the slab's dipole moment
to the dielectric's macroscopic properties.
Next, we address the problem of an adequate boundary condition at the bottom of
the slab  which could pin the chemical potential within the SiC gap regardless 
of the specific polytype. To this end we test various 
bottom terminations, including a H capping layer with and without an additional 
graphene layer as well as an ultrathin gold film of different thicknesses, to find that 
the latter solves satisfactorily the chemical potential problem. Semi-infinite QFG 
surfaces are next constructed and solved via Green's function methods in order to 
attain an accurate description of the Dirac cones and allow a precise estimation of 
their doping for each SiC polytype. Our scheme also provides a detailed picture of the 
development of the SP as a function of the slab thickness which should be relevant in 
the context of QFG on thin SiC films. We will assume throughout a perfect defect-free 
graphene layer thus restricting the study to SP induced dopings. The combined effect of
SP and self-doping, on the other hand, is even more challenging and will be presented 
in a separate work~\cite{letter}.

The paper is organized as follows: in Section II the theoretical details of the DFT 
calculations are summarized. In Section III we present the slab formalism 
employed to study the bulk SiC dielectric properties together with the derived
values for the relative permittivities and the SPs.
The central findings of the paper are given in Section IV, where (i) different
boundary conditions at the bottom are addressed, (ii) the QFG's doping due to the 
substrate's SP is calculated as a function of the slab thickness and, (iii) a 
simple electrostatic model based on macroscopic averages of the DFT results
is analyzed in order to account for the calculated dopings. The last section
is devoted to a final discussion and the conclusions.

\section{Theoretical details}
\begin{figure}
 \includegraphics[scale=0.35]{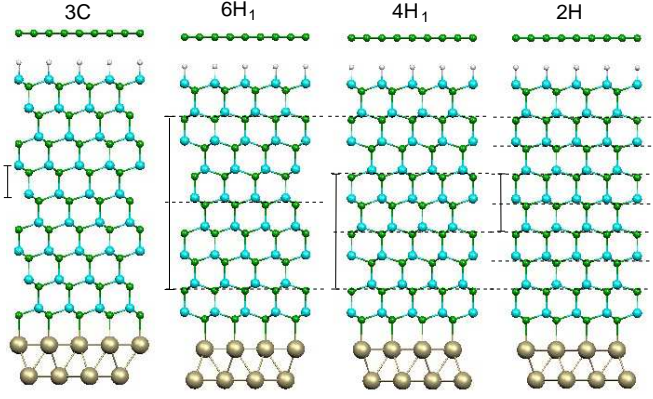}
\caption{Geometries of the 9~BLs thick G/H/(SiC)$_9$/Au$_2$ slabs employed to 
         model the QFG for different SiC polytypes. The two planes at the bottom
         correspond to the capping Au layer (see Section IV.A). 
         The out of plane lattice parameter $c$, is indicated by vertical solid
         lines while dashed horizontal lines are
         drawn at the location of the SDs. The subindexes in 6H$_1$ and 4H$_1$
         indicate that the first SD is the closest possible to the surface.
         \label{geom} }
\end{figure}

All DFT calculations have been performed with the pseudopotential SIESTA formalism~\cite{siesta} (as implemented within the GREEN package~\cite{green}) and under the generalized gradient approximation (GGA).\cite{gga} We generated the atomic orbital basis set according to the double-$\zeta$ polarized (DZP) scheme employing confinement energies of 200~meV for all elements. Real space meshes with a resolution of $\sim$0.06~\AA$^3$ (mesh cut-off set to 700~Ryd) were defined for performing the 3-center integrals. Unless otherwise stated, the temperature used in the Fermi-Dirac distribution was $k_B$T=10~meV while dipole-dipole interactions among neighbor supercells were supressed via the usual dipole-dipole
corrections.\cite{dipole} Figure~\ref{geom} shows the geometries of all considered polytypes, cubic 3C-SiC(111) as well as hexagonal 2H-, 4H- and 6H-SiC(0001) in a slab geometry. The density of SDs increases with the hexagonality of the polytype, presenting a SD every four, three and two bilayers (BLs) in the 6H, 4H and 2H structures, respectively. The corresponding lattice constant along the surface normal, $c$, contains two SDs and therefore, is six, four and two times longer than that of the cubic 3C (vertical solid lines in the figure). In all calculations the in-plane lattice constant was always fixed to the experimental value of $a^{\mathrm{exp}}$=3.08~\AA\ (our GGA optimized value is 3.10~\AA), while $c$ was optimized for 
each polytype leading to the values given in Table~\ref{table}. The SDs only cause marginal expansions (below 0.5~\%) of the inter-BL spacings.

Two types of 2D slabs have been considered. First, and in order to address the SiC dielectric properties, we defined (1$\times$1) slabs, H/(SiC)$_n$/H, oriented along the (111) (or (0001) for polytypes) direction of different thicknesses $n$, with both the top (Si) and bottom (C) layers fully hydrogenated. A $k$-sampling of (30$\times$30) for all these (1$\times$1) slabs was found enough to achieve well converged values of the SP. 

For the second set of slab calculations we added a graphene layer on top of the upper H capping layer, G/H/(SiC)$_n$/X, and assumed a simplified $(2\times 2)/(\sqrt{3}\times\sqrt{3})R30^\circ$ commensurability between the G/SiC(111) lattices with the lattice constant of the slab set to that of the SiC, thus forcing an 8\% expansion of the C-C bonds in the graphene with respect to that in the $(13\times 13)/(6\sqrt{3}\times 6\sqrt{3})R30^\circ$ Moir\'e pattern experimentally 
observed~\cite{goler} (this should, nevertheless, have a minor impact on the calculated doping values). As detailed in section IV.A, different terminations X at the bottom of the slab were considered. We included van der Waals dispersion forces for a proper description of the graphene-substrate interactions following the semiempirical scheme of Ortmann and Bechstedt.\cite{vdw,ourcarbon} In the total energy optimizations we relaxed the first upper and lower surface layers in the slabs until forces were smaller than 0.02~eV/\AA\ while the geometry of the most internal SiC BLs was set to that optimized for the corresponding bulk polytype.

\begin{figure}
\includegraphics[scale=0.45]{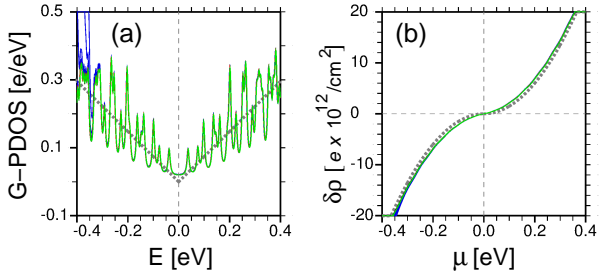}
\caption{(a) DOS projected on the graphene layer. Results for the G/H/(SiC)$_n$/Au$_2$  
slabs with $n=8-12$ and all polytypes are superimposed in different colors (see Fig.~\ref{1x1}).
The energy origin for each case is located at its DP while the broadening was set to 5~meV.
(b) Charge doping of the graphene layer as a function of the chemical potential
$\mu$ with respect to the DP and for the same cases as in (a).
Thick dashed lines in (a) and (b) show the best linear fit to the DOS 
(obtained for a reduced Fermi velocity of $v_F$=0.7~m/s)
and the associated quadratic surface charge density, respectively.
         \label{qg} }
\end{figure}

We highlight the fact that, in order to obtain well converged results for the 
(2$\times$2) slabs, specially regarding the QFG doping charges, we required unusually 
fine reciprocal space $k$-meshes as large as (100$\times$100) in the self-consistent 
Hamiltonian's calculations thus increasing considerably the computational effort.
In Figure~\ref{qg}(a) and (b) we plot the graphene projected density of states
(PDOS) and its associated charge, respectively. For a broadening of 5~meV the
PDOS presents a spiky structure but, fortunately, the $e$ and hole charges are
well converged and can be accurately fitted by a quadratic energy dependence 
with a Fermi velocity of $v_F$=0.7~m/s (for an isolated graphene layer we 
obtain $v_F$=0.9~m/s, which is about 20\% smaller than the experimental value).

Once self-consistent Hamiltonians were obtained for the QFG slabs, we constructed true semi-infinite surfaces after replacing the bottom layers of the slabs by a semi-inifinte bulk following the Green's functions based prescription detailed elsewhere.\cite{loit} Band structures in the form of $k$-resolved density of 
states, PDOS($k,E$), projected on the QFG, H and uppermost SiC BLs were 
evaluated employing a broadening (imaginary part of energy) of 5 meV. 

\section{Bulk SiC dielectric properties}
In this section we address the bulk dielectric properties of different SiC 
polytypes, namely: 3C-SiC(111) and 2H-, 4H- and 6H-SiC(0001). Although different
schemes have been proposed to study these properties, most of them relying on 3D
unit cell calculations~\cite{vanderbildt,sp2physicab,qteish,sp1},
here we will adopt an alternative approach based on 2D slabs of different 
thicknesses~\cite{sr,fu} given its simplicity and because
the impact of the SP at surfaces will be studied under this model geometry. 
We will revisit below the slab 
formalism in terms of the slab's dipole moment and its relationship to the
{\it macroscopic} bulk polarization. Despite the derivation of such
expression being based on standard electrostatic theory, we find it appropiate
to present the entire formalism in detail since, after inspecting a vast number 
of works in the field, we were not able to find a general and explicit equation
analogous to the one derived below. 
In appendix A we present a parallel study based on the bulk formalism
of Qteish~{\it et al}~\cite{qteish}, which we find less precise.

\begin{figure}[!hbt]
\includegraphics[scale=0.55]{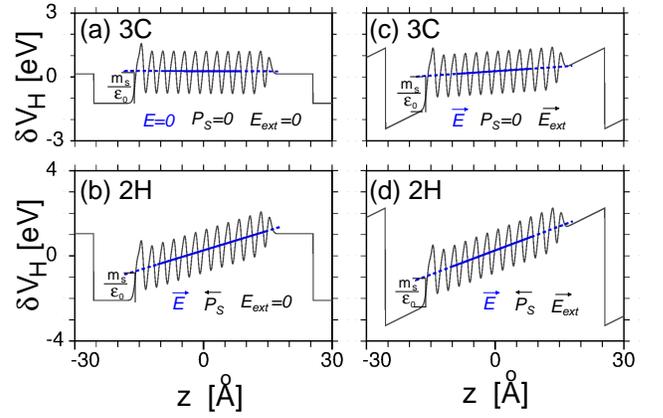}
\caption{$\delta V_H(z)$ profiles after averaging over the 2D unit cell for 
         H/(SiC)$_{12}$/H slabs calculated under DDCs boundary conditions
         assuming a (a) 3C-SiC(111) and (b) 2H-SiC(0001) substrate. 
         (c) and (d) are the same as (a) and (b),
         respectively, after applying an external field $\Ee$=0.1~eV/\AA.
         In each graph thick blue lines in the inner region of the dielectric
         correspond to macroscopic averages, $\vhav{z}$, obtained via gaussian 
         smearing employing a width of 5~\AA.
         The potential discontinuity at the vacuum region, $\Delta V$,
         appears at the right and left of each plot.
         The DDCs leave a localized dipole at the left surface
         (H-C termination) responsible for the $m_s/\epso$ potential drop at
         the vacuum-surface interface, indicated in each plot 
         after  extrapolating $\vhav{z}$ (thick dashed blue lines).
         On the other hand, the dipole and potential drop at the right surface
         (H-Si termination) is almost negligible.
\label{macro-1x1}}
\end{figure}
Throughout this section we consider (1$\times$1) H/(SiC)$_n$/H slabs with 
$n=6-12$ the number of SiC bilayers (BLs) while hydrogen capping layers are
adsorbed both at the top (Si terminated) and bottom (C terminated) surfaces in 
order to saturate the dangling bonds and reduce the slab's dipole moment
arising from the polar character of the SiC(0001) surface.
As pointed out in Section II, all calculations include the usual dipole-dipole 
corrections (DDC) among image slabs so that the electric field in the vacuum 
arising from the slab's dipole is supressed\cite{dipole}. 
In Figure~\ref{macro-1x1} we present the planar averaged Hartree potential~\footnote{Throughout this work we will refer to $\delta \rho$ as
the total charge density minus that of the isolated atoms while 
$\delta V_H$ to its associated electrostatic 
potential. We show these quantities rather than the total ones, 
$\rho$ and $V_H$, since the formers provide a better visual resolution.
On the other hand, macroscopic averages of their profiles across the slab
are obtained via gaussian smearing employing widths of 2-6~\AA.}
profile along the slab's normal, $\delta V_H(z)$, for the 12~BLs thick 3C and 2H
slabs without and with an applied external field ($\Ee$=0.1~eV/\AA). 
The corresponding macroscopic average $\vhav{z}$ within the 
dielectric is superimposed as a thick blue line, from which the local electric 
field may be extracted via $\Eav=-\partial{\vhav{z}}/\partial z$ (recall 
$\delta V_H(z)$ is given in $-|e|V$ units).

In order to establish the connection between the macroscopic bulk dielectric
properties we are seeking and the quantities that may be extracted from the
DFT calculations we express the slab's dipole moment per unit area, $m$,
in terms of the polarization $P$ in the dielectric 
and its thickness $t$ as:
\begin{equation} \label{eq:m1}
 m(t) = m_s + P\; t
\end{equation} 
where, under DDC boundary conditions, the $m_s$ constant is a surface
localized contribution that accounts for the polar
character of the SiC slab and, as will be shown below, is only slightly
dependent on the actual SiC polytype.
In the absence of polarization $m_s$ determines the 
potential drop at the vacuum region: $\Delta V= m_s/\epso$ (Fig.~\ref{macro-1x1}(a)),
while for the general $P\neq0$ case, $\Delta V$ has additional
contributions from the external field and/or the SP (Figs.~\ref{macro-1x1}(b-d)).

In the above equation $P$ represents the macroscopically averaged
dipole moment density in the dielectric assuming it is
homogeneous --that is, does not depend on $z$. In general, $P$
may be split into two terms: 
\begin{equation} \label{eq:m2}
 P = P_S + P_E = P_S + \epso \chi_e \Eav
\end{equation}
where $P_S$ corresponds to the SP strictly defined as the polarization 
present in the slab at zero local field ($\Eav=0$) while
$P_E$ is the polarization induced by the presence of a finite
local field ($\Eav\neq 0)$
--for the last equality we
have further assumed that the dielectric is linear with a
susceptibility $\chi_e=\epsr-1$. 

Applying the continuity equation for the displacement field across any of the 
two vacuum-dielectric interfaces we arrive at:
\begin{equation} \label{eq:m3}
\epso \epsr \Eav + P_S = \epso \Ee
\end{equation}
Thus, even if $\Ee=0$, a finite value of $P_S$ will be associated
to a non-vanishing local field $\Eav$ inside the slab 
(see Fig.~\ref{macro-1x1}(b)).

Combining eqs.\eqs{m2}{m3} we arrive at a general expression
for the total polarization which 
does not depend on the macroscopically averaged field $\Eav$:
\begin{equation} \label{eq:m4}
 P = \frac{P_S}{\epsr} + \frac{\chi_e}{\epsr}\epso\Ee
\end{equation}
Comparison of eq.\eq{m2} versus\eq{m4} clearly establishes that
the SP defined at zero local field and that defined at zero
external field differ by a $1/\epsr$ factor~\cite{fu}. It also follows
that if $P_S\neq0$, the $\Ee$ contribution in the second term above
should not be identified with the induced 
polarization $P_E=\epso\chi_e\Eav$.

Finally, inserting eq.\eq{m4} in\eq{m1} we arrive at our desired
expression for $m(t)$ valid under DDCs boundary conditions:
\begin{equation} \label{eq:m5}
 m = m_s + \left( P_S + \chi_e\epso\Ee \right) \; \frac{t}{\epsr}
\end{equation}

To our knowledge, and despite its simplicity, eq.\eq{m5} has not been
explicitly reported before. For instance, in Ref.[\onlinecite{fu}], where the 
SP for BaTiO$_3$ was studied also employing slab geometries, the $1/\epsr$
factor was included via somewhat heuristic arguments. 

In the spirit of SR, the unknowns in eq.\eq{m5}, namely $\epsr$, $P_S$
and $m_s$, may be extracted after fits of $m(t)$ curves calculated for
different slab thicknesses and/or external fields. This is already an
advantage versus bulk approaches where $\epsr$ and $P_S$ are typically obtained
from independent calculations, or even the experimental value of $\epsr$
is employed\cite{qteish}. Below, we will derive first $\epsr$ and next $P_S$
from eq.\eq{m5} following a two stage linear fitting scheme --we
found this approach more accurate than performing a simultaneous non-linear fit
for both unknowns.


\subsection{Relative Permittivities}
The calculation of the relative permittivites, $\epsr$, for the various SiC 
polytypes is straightforward and may be regarded as a benchmark to test the 
accuracy of the calculation parameters described in the previous section. 
Taking the partial derivative with respect to $t$ in eq.~\eq{m5}, we have:
\begin{equation} \label{eq:epsr}
\epsilon_r = \frac{\epsilon_0 \Ee - P_S}{\epsilon_0 \Ee - \partial m/\partial t}
\end{equation}
The existence of a finite SP ($P_S\neq0$),
however, requires eq.\eq{m5} to be fitted with certain care. Here, we
perform two sets of calculations for each polytype, one under a positive 
external field, $\Ee=0.1$~eV/\AA\ and a second one under a negative field,
$\Ee=-0.1$~eV/\AA. Denoting by $m^\pm$ the dipole moment under $\pm \Ee$, we
may substract them to eliminate the $P_S$ and $m_s$ contributions in 
eq.\eq{m5} to obtain $\delta{m}=\frac{1}{2}\left({m^+-m^-}\right)=
\frac{\chi_e}{\epsr}\epso\Ee\; t$, so that the relative permittivity may be
directly calculated from the $\partial \delta m/\partial t$ slope via:
\begin{equation} \label{eq:epsr2}
\epsilon_r = \frac{\epsilon_0 \Ee}{\epsilon_0 \Ee - \partial \delta m/\partial t}
\end{equation}
If the atoms are relaxed under the presence of the electric field eq.\eq{epsr2} 
provides the static permittivity $\epsilon_r(0)$, while if they are fixed to 
their zero field equilibrium positions it gives the high-frequency permittivity,
$\epsilon_r(\infty)$~\cite{sr}.

\begin{figure}
\includegraphics[scale=0.45]{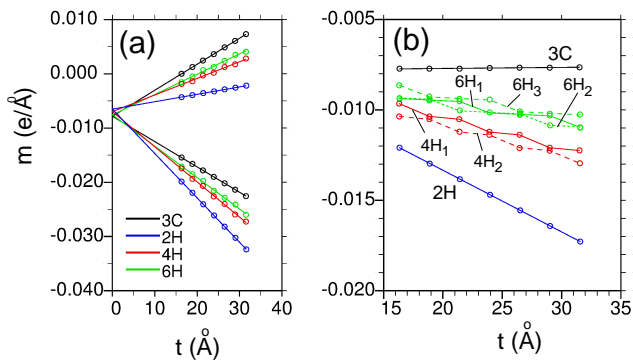}
\caption{DFT derived bulk dielectric properties for SiC. (a) Dependence of the 
 dipole moment per unit area, $m$, with the H/(SiC)$_n$/H slab thickness, $t$, 
 for 3C-, 6H$_1$-, 4H$_1$- and 2H-SiC polytypes. The slabs thickness ranges
 between 6 and 12 BLs. Positive and negative curves 
 calculated under external fields of +0.1~eV/\AA\ and -0.1~eV/\AA, respectively.
 Open circles correspond to the calculated dipole moments with fixed
 geometry  while the straight lines are best fits to the data points.  
 (b) Same as (a) but in the absence of an external electric
 field. Solid lines link the fitted values using approximation~\eq{P0}
 for $P_S$ in eq.~\eq{m5}.
 Blue, red, green and dark lines correspond to the
 2H, 4H$_{1/2}$, 6H$_{1/2/3}$ and 3C cases, respectively (see text for further
 explanations). 
 \label{1x1}}
\end{figure}

In Fig.~\ref{1x1}(a) we show the $m(t)$ dependence for (1$\times$1) H/(SiC)$_n$/H slabs 
with $n=6-12$ and for all polytypes under $\Ee$=$\pm$0.1~V/\AA\ keeping fixed 
the geometry. The correct performance of the approach can be judged by the 
almost perfect linear behavior in all plots and the resulting high frequency 
permittivities $\epsilon_r(\infty)=7.0-7.3$ (see Table~\ref{table}), 
only slightly larger than the reported experimental values for SiC at room 
temperature of $\epsilon_r(\infty)=6.5$\cite{sic_exp}. 
Similar linear plots are obtained if the
slabs' geometries are relaxed (not shown), yielding increased static
permittivities of $\epsilon_r(0)=10.3-10.9$ (Table~\ref{table}) which are again
in good agreement with the experimental value of 
$\epsilon_r(0)$=9.7.\cite{sic_exp}
The small differences among polytypes
arise from the fact that $P_S$ and $\epsr$ are coupled in eq.\eq{m5}. 
On the other hand, we find the expected trend that as the hexagonality of the 
dielectric increases, the asymmetry between the $\partial m^\pm /\partial t$ 
slopes also increases due to the larger SP; that is,
the response of the dielectric to an external field will be
different if it already shows a finite polarization at zero
field.

Finally, the intercept of $m(t)$ with the ordinate axis provides the DDC dipole
moment $m_s$ arising from the polar character of the SiC slabs (which is substantially
reduced, but not fully removed, by the capping H layers). In our case, $m_s$
attains essentially the same negative value for the 3C, 4H and 6H polytypes, 
while the 2H slightly deviates towards a smaller absolute value since the
SP partially counteracts $m_s$. 

\begin{table}
\caption{Calculated lattice parameter $c$ in \AA, 
         dynamic and static relative permittivities, 
         $\epsilon_r(\infty)$ and $\epsilon_r(0)$, respectively, 
         and the spontaneous polarization values, $P_S$, in \sigmauC
         (\sigmau),
         for the different SiC polytypes considered in this work.
         \label{table}}
\begin{tabular}{ccccc}
\hline \hline
    &     $c$     &   $\epsilon_r(\infty)$ & $\epsilon_r(0)$ & $P_S$  \\
\hline
 3C &    2.53     &            6.96        &       10.3      &+0.1 (+0.4) \\
 6H &   15.22     &            7.26        &       10.8      &-1.2 (-7.5) \\
 4H &   10.14     &            7.32        &       10.5      &-2.0 (-12.5) \\
 2H &    5.08     &            7.35        &       10.9      &-4.0 (-25.0) \\
\hline \hline
\end{tabular}
\end{table}

\subsection{Spontaneous Polarizations}
Once $\epsr$ is known, eq.\eq{m5} may be applied to the same set of $\pm\Ee$ 
slab calculations in order to estimate the SP of each SiC polytype via:
\begin{equation}
P_S= \epsr \frac{1}{2}\; \frac{\partial(m^+ + m^-)}{\partial t}
\end{equation}
Instead, we present below the SP analysis based on similar slab calculations 
but in the absence of an external field ($\Ee=0$) in which case $P_S$ takes
the even simpler form $P_S= \epsr \; \frac{\partial m}{\partial t}$.
Hereafter we also pay attention to the location
of the first SD relative to the top Si layer and arrange the 6H (4H) slabs into
three (two) subsets, 6H$_{1-3}$ (4H$_{1-2}$), where the subindex increases as 
the first SD is located further away from the uppermost BL. 

In Fig.~\ref{1x1}(b) we
plot the dipole moment per unit area for all H/(SiC)$_n$/H systems with 
$n$=6$-$12. 
For the SD free SiC(111)-3C slab (black line) no SP exists
and, according to eq.\eq{m5}, the dipole moment (per unit area) should remain 
fixed to $m_s$ independent of $n$,
as it is indeed the case (the slope yields a negligible SP of
$P_S^{3C}=7$~C$\times$10$^{-4}$/m$^2$). 
On the other hand, the dipole moment for the 2H slabs 
(blue line) presents an almost perfect linear dependence with $t$ due to the 
absence of crystalline regions, yielding a value of $P_S^{2H}=-4.0$~\sigmauC\ 
which is in reasonable agreement with previous works given the large
scatter among the reported values ($P_S^{2H}=-1.1$ to -4.3~\sigmauC)~\cite{sp2physicab, qteish, troschin}.

The 4H and 6H cases (red and green lines, respectively) show a somewhat
different behavior and vary in a non-linear way with $n$. They present
rather flat slopes with sudden drops whenever an additional SD is incorporated 
in the slab. The drops always attain similar values regardless of the slab 
thickness, the location of the first SD or the actual polytype (see below). 
Hence, they may be identified with the dipole moment per unit area
associated to a single SD, $m_{SD}$, while
the horizontal sections correspond to the SP in the crystalline regions of the 
slab, $P_{S,c}$. In fact, they are not strictly flat, but present 
negative slopes with associated SP values that decrease in absolute value
with the hexagonality,
$P_{S,c}^{4H_{1/2}}=-0.7$~C$\times$10$^{-2}$/m$^2$ and
$P_{S,c}^{6H_{1/2/3}}=-0.4$~C$\times$10$^{-2}$/m$^2$, implying that the wider
the crystalline region, the more efficiently is $m_{SD}$ screened. 
Notably, and as indicated by the lines in the figure, within each subset the 
dipole moments can be very accurately fitted by setting:
\begin{equation}\label{eq:P0}
  P_S(t)= P_{S,c} + \epsilon_r m_{SD} N_{SD}(t)/t
\end{equation}
in eq.\eq{m5}, where $N_{SD}(t)$ is the number of SDs in the slab. 
The bulk SP for a given polytype is then simply given by:
\begin{equation}\label{eq:PS}
 P_S \approx \frac{\partial (P_S(t) \; t)}{\partial t}= P_{S,c} + 2 \epsr m_{SD}/c
\end{equation}
where $c$ is the length of the bulk repeat vector along the slab's normal 
(specific of each polytype and given in Table~\ref{table}) and the factor two
accounts for the fact that there are two SDs per repeat unit 
(see Fig.~\ref{geom}). The fits employing eq.\eq{PS} yield a dipole moment
per SD of $m_{SD}=5.5\times10^{-4}e$/\AA\ and SP
values of $P_S^{4H_{1/2}}=-2.0$~\sigmauC\ and 
$P_S^{6H_{1/2/3}}=-1.2$~\sigmauC, again, in close agreement with 
previous estimates based on bulk 
calculations~\cite{sp2physicab,qteish,troschin,brandino} (see also Appendix A).

It also follows from eq.\eq{P0} and Fig.~\ref{1x1}(b) that if within each 
subset of slabs we only consider in the fits those whose thicknesses differ by 
$c/2$ (that is, $\delta n=2m$ for the 4H and $\delta n=3m$ for the 6H cases, 
with $m$ integer) then $P_S$ becomes independent of $t$ and may be directly
obtained from the resulting linear slopes.

\subsubsection{SP dependence with $e$/hole concentration}
\begin{figure}
 \includegraphics[scale=0.4]{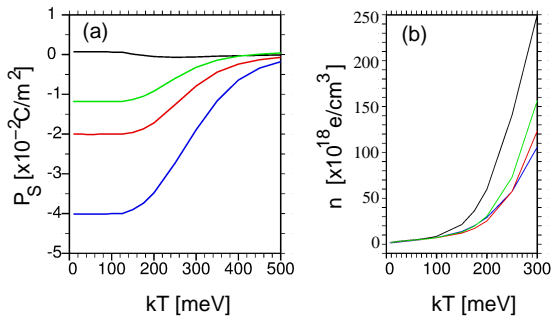}
\caption{Evolution of (a) the SP and (b) the electron/hole concentration
         as a function of the {\it electronic} (or Fermi-Dirac) 
         temperature, $T$, for all polytypes (same color scheme as in 
         Fig.~\ref{1x1}). In (b) we only show results for
         the 12~BLs thick slabs.
\label{sp-kT}}
\end{figure}

Last, we explore the robustness of the SP versus the $e$/hole concentration in the 
{\it intrinsic} dielectric. To this end, we have recalculated the electronic 
structure of 
all H/(SiC)$_n$/H slabs self-consistently at different temperatures, $T$. 
More precisely, and since $T$ only enters our calculations when computing the 
occupation of states via the Fermi-Dirac distribution function, we estimate the
evolution of $P_S$ with the 
density of bulk {\it free} charges or, equivalently, with the hole/electron 
concentration at the valence/conduction bands (we disregard, however, the 
dependence of $\epsr$ with $T$). 
The results are displayed in Fig.~\ref{sp-kT} where we find the 
expected decrease of the SP in (a) as the electron/hole concentration
shown in (b) increases. 
At typical SiC doping concentrations ($\sim 10^{18} e/$cm$^3$) all
$P_S$ values remain constant while beyond threshold concentrations of 
$\sim 10^{19} e/$cm$^3$ (or $k_BT>120$~meV) they start to decrease rapidly due 
to the screening of the internal dipole moments by the {\it bulk free} charges. 
Obviously, the threshold free charge concentration increases as the SP of the 
polytype does. Since the equivalent {\it electronic} temperature threshold is 
far above RT ($T>1000$~K), the SP in SiC(0001) samples may thus be considered
as highly robust versus temperature or bulk dopings.

\section{Relationship between the doping of graphene and the SP}
We now focus on the main point in this work, which is the estimation, from first principles calculations, of the doping of the Dirac cones in the QFG surface system due to the SP.

\subsection{Slab models for QFG}
\begin{figure*}
\includegraphics[scale=0.75]{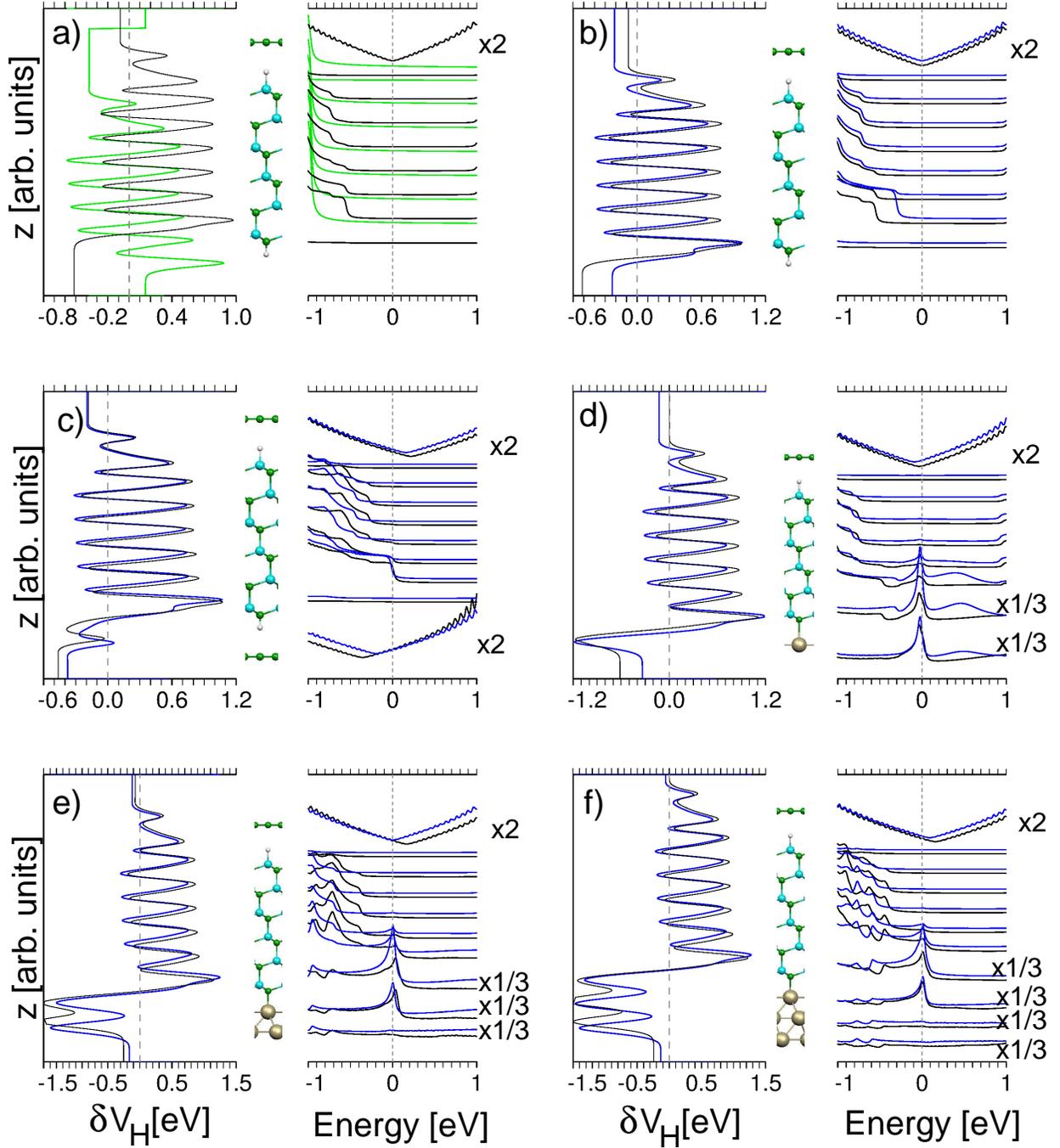}
\caption{(a) left: $\delta V_H(z)$ profile along a G/H/(SiC)$_6$/H slab depicted at the
         center and, right: the corresponding DOS projected, in ascending order, on the bottom H layer, the six SiC BLs, the intercalated H layer and the graphene; green (dark) lines correspond to a slab calculation without (with) the H capping layer at the bottom.
         (b) Dark lines same as in (a) while blue lines correspond to the same 
         H-saturated slab after elongating the C-H bonds at the bottom by 0.45~\AA.
         (c) Same as (b) but for a G/H/(SiC)$_6$/H/G slab; dark lines correspond to the relaxed geometry and blue after elongating the C-H bonds by 0.4~\AA\ and shifting the bottom graphene layer by another 0.4~\AA.
         (d) Same as (c) but for a G/H/(SiC)$_6$/Au slab; dark lines for the relaxed geometry and blue after expanding the Au-C
         bonds by 0.5~\AA. 
		 (e) Same as (d) but for a G/H/(SiC)$_6$/Au$_2$ slab; blue lines correspond to expansion of the Au-C bonds and the Au
         interlayer distances by 0.5~\AA\ and 0.65~\AA, respectively, with respect to the relaxed geometry (dark). 
		 (f) Same as (e) but for a G/H/(SiC)$_6$/Au$_3$ slab after applying 0.4~\AA\ elongations to the C-Au and both Au-Au interlayer spacings (blue) with respect to the relaxed geometry (dark).
         \label{au} }
\end{figure*}

Let us first address in detail the drawbacks of the slab geometry when modeling
polar surfaces. To this end, we consider the QFG system G/H/(SiC-3C)$_6$/X
already described in Section~II, and examine different terminations X at the 
bottom of the slab in order to reduce the surface dipole and obtain 
a boundary condition at the bottom of the slab which could reasonably mimic that
expected from a semi-infinite SiC(111) surface. Essentially, we look for 
electronic states within the gap and localized at the bottom of the slab which 
could lead to a well defined chemical potential, $\mu$, in a similar manner as 
dopant impurities determine the chemical potential in a real dielectric. 
Although the discussion below is restricted to six BLs thick slabs, we have 
checked in all cases that increasing the slab thickness up to twelve BLs does 
not alter our conclusions.

We start with the most common practice of saturating the C dangling bonds at the
bottom of the slab with H atoms. In Fig.~\ref{au}(a) we plot the Hartree 
potential profile, $V_H(z)$, before (green line) and after (dark) adding the H 
capping layer. A reversal of sign and a substantial decrease of the surface 
dipole, $m_s$, is immediately obvious from the reduction of the potential step between 
the vacuum regions at both sides of the slab.
In the same figure we present the graphene and BL resolved
DOS for both cases. For the slab without H atoms at the bottom no trace of the 
graphene bands is seen in the energy window due to the huge doping induced by 
the unsaturated C atoms. Under the presence of the H-capping layer, on the other
hand, the chemical potential (or Fermi level) remains fixed at the DP and within
the gap. The surface dipole may be reduced by expanding the C-H bond lengths at
the bottom of the slab thus generating a local dipole that may counterbalance 
the former. The resulting potential after an outwards 0.45~\AA\ displacement of
the saturating Hs is shown by the blue lines in Fig.~\ref{au}(b). Although the 
potential step is essentially removed, the position of the DP remains pinned at
$\mu$. This is a consequence of the absence of gap states at the bottom of the 
slab, so that charge neutrality forces the bands of graphene to follow any band
bending (BB) and pins $\mu$ at the DP. Indeed, as long as the BB does not cross
into the conduction or valence bands this picture will remain regardless of the
presence of any internal dipoles in the slab. Therefore, slab models with a H 
capping layer at the bottom are not suitable for the estimation of any 
SP-derived doping.

A natural way of introducing gap states could be to add another graphene layer at the bottom leading to a more symmetric G/H/(SiC)$_n$/H/G geometry. 
The Hartree potential and DOS for such case are given by the dark lines in Fig.~\ref{au}(c). We now find an enhanced BB which leads to the pinning of $\mu$ at the valence band edge of the lower SiC BL together with large $n$- and $p$-type dopings at the bottom and top graphene layers, respectively, that compensate
each other. 
Expanding the bond lengths at the bottom of the slab by considerable amounts (blue lines) hardly changes the doping level at the top graphene and hence, this model can also be ruled out for the estimation of any influence of SPs.

Our next trial model consists of replacing the H capping layer by a metallic one
with the hope that the creation of a metal-induced gap states (MIGS) could 
effectively pin $\mu$ within the gap. As shown in Fig.~\ref{au}(d)-(f), this is
indeed the case when one, two or three Au layers, respectively, are used to 
passivate the C dangling bonds. We found energetically more favorable to place 
the Au layer in contact with the C atoms at top positions while additional Au 
layers are stacked following an $fcc$ sequence. For the 1~ML case, dark line in
(d), the MIGS appears as a large peak in the middle of the gap which penetrates
up to three bilayers into the dielectric. The top graphene is now only slightly
$n$-doped and although the dipole is still considerable it may be again 
suppresed by expanding the Au-C spacings (blue lines). The expansion leads to an
enhancement of the MIGS's DOS and a slightly larger doping. Adding a second Au 
layer changes the doping to $p$-type with the MIGS still pinning the chemical 
potential within the gap although this time slightly closer to the valence band
(dark lines in (e)). In order to compensate the surface dipole we now require 
Au-C and Au-Au expansions as large as 0.50 and 0.65~\AA, respectively, but with
the fortunate outcome of removing the doping and leaving the DP aligned with 
$\mu$. For the sake of completeness, we present in (f) the case of three Au 
layers, where a moderate $p$-doping is now obtained even after elongating the 
interlayer spacings.

In summary, we find that $\mu$ can be pinned at the bottom of the slab and 
within the gap after capping the C atoms with an Au layer via the appearance of
a MIGS. 
Furthermore, the surface dipole can be removed by expanding the Au-C and
Au-Au interlayer distances by large amounts ($\sim 0.5$~\AA) while the 
actual doping in the graphene layer can be tuned by choosing the thickness of 
the metallic layer, obtaining $n$-doping for one atomic plane, no doping for two
and $p$-doping for three. 
The pinning of $\mu$ occurs due to the much larger DOS of the MIGS
compared to that of the QFG, and we stress that it is an essential prerequisite
to make meaningful any differences in the graphene doping among different
polytypes. Also note that our model slabs may be as well employed to quantify 
any doping in the graphene arising from defects~\cite{letter}.

\subsection{Doping of graphene due to the substrate's SP}
\begin{figure*}[!hbt]
\includegraphics[width=2.0\columnwidth]{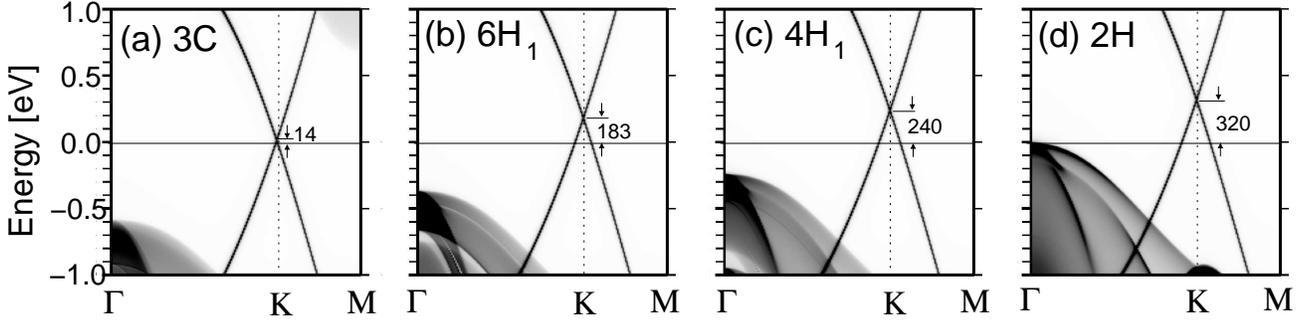}
\caption{DOS($k, E$) projected on graphene, H and first three SiC BLs of a G/H/SiC semi-infinite surface calculated after matching the Green's function of a G/H/(SiC)$_{12}$/Au$_2$ slab with that of the corresponding bulk (see Sec. II for further details). Four different SiC polytype surfaces are shown: (a) 3C, 
  (b) 6H$_1$, (c) 4H$_1$ and (d) 2H. The DP shift,  $\Delta$DP, is indicated in
  each plot in meV. 
                  \label{dosk}}
\end{figure*}

\begin{figure}[!h]
\includegraphics[scale=0.70]{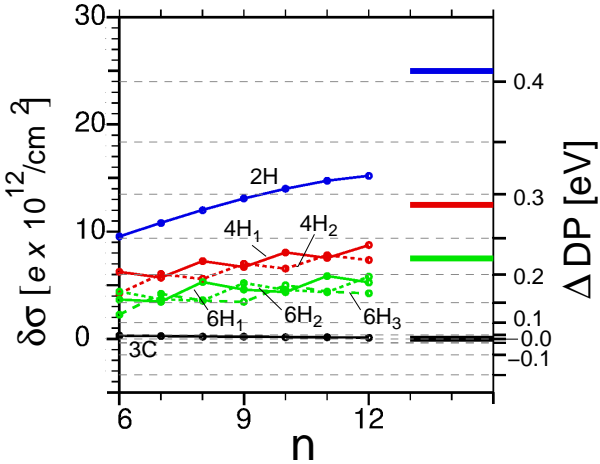}
\caption{Doping of the graphene layer for all G/(SiC)$_n$/Au$_2$ slabs 
         considered in this work as a function of $n$, the SiC polytype and the
         location of the SD closest to the surface (same color scheme as in 
         Fig.~\ref{1x1}); left axis gives the surface charge density and right 
         axis (quadratic scale) the Dirac point shift, $\Delta$DP=DP$-\mu$. 
         Horizontal lines at the right give the bulk SP 
         associated to each polytype (see Table~\ref{table}).
\label{dope}}
\end{figure}

Once we have proven that both the SP and the QFG doping can been evaluated 
under the same slab-based framework we may explore their interplay
as a function of the SiC polytype and the slab thickness.
We choose the G/H/(SiC)$_n$/Au$_2$ slab with elongated bonds at the bottom
shown in Fig.~\ref{geom} as
our model system for all calculations presented in this section since 
in the absence of SDs it yields hardly any doping (see Fig.~\ref{au}(e)).
Alternatively one may use the Au$_1$ termination if a small $n$-type doping
is desired for the 3C case (as often found experimentally --see Table~\ref{alldope})
or the Au$_3$ termination for a mild initial $p$-type doping.
Both terminations should anyhow yield similar dopings
if the DP location for the 3C case is used as the reference when comparing
against the rest of polytypes.

In order to achieve a more accurate picture of the surface electronic structure
we calculated the graphene and SiC projected density of states (PDOS) under
a semi-infinite geometry after replacing the Hamiltonian matrix elements 
involving the lower layers in the 12~BL thick slab by those corresponding to an
ideal bulk termination as outlined in Section~II~\cite{loit}. 
Figure~\ref{dosk} shows $k$-resolved PDOS on graphene, the intercalated H layer
and the first three SiC BLs for the 3C, 6H$_1$, 4H$_1$ and 2H slabs for
the maximum thickness considered, $n$=12. 
The semi-infinite
geometry provides a continuum of states for the valence band while
the Dirac cones across the gap are clearly visible in all plots. 
In accordance with the experimental trend, the DP shift with 
respect to the chemical potential, $\Delta$DP, increases with the hexagonality of 
the polytype; starting from a marginal value of 14~meV for the SD free surface 
(a), we obtain a value as large as $\Delta$DP=320~meV for the 2H case (d), 
that is, equivalent to a $p$-type doping surface charge density of 
$\delta\sigma =$ 17~\sigmau. The 6H$_1$ and 4H$_1$ surfaces (b)-(c), also show 
substantial shifts of 183 and 240~meV, respectively, corresponding to charges 
in the $5-10$~\sigmau\ range.

It is important to note, however, that despite the model system being
semi-infinite, the doping
calculated for each polytype depends on the particular slab 
employed to perform the matching with the bulk. This is because the Fermi level
and the Hamiltonian matrix elements employed for the surfacemost layers in
the semi-infinite are extracted from the slab calculation and, hence, they 
implicitly contain the DP shift, whereas those employed for the bulk like layers
are extracted from a separate bulk calculation in which no band bending can 
occur due to the periodic boundary conditions (in fact, the $\Delta$DP values
differ by less than 5~meV when deduced from equivalent PDOS plots extracted 
directly from the slab calculation). 

Fig.~\ref{dope} shows the calculated graphene doping
for all surface systems as a function of the number of BLs, 
$n$, included in the slab. 
For the 6H and 4H stackings we again take care of
the location of the uppermost SD and group the results accordingly (see III.B). 
For the 3C case the DP remains close to $\mu$ for all thicknesses, thus 
corroborating the general validity of our
slab model as no doping is expected in the absence of SP.
On the other hand, the correlation between $\delta\sigma$ and the SP becomes 
patent after noting the stairlike behavior of the doping for the 4H and 6H 
polytypes, highly reminiscent of that appearing in Fig.~\ref{1x1}(b).
The dopings depend on the number of SDs in the slab, with positive jumps 
whenever a new SD is incorporated while if the added BL follows the
cubic stacking the doping decreases only slightly.
Overall, the surface charge densities, $\delta\sigma$, increase almost
linearly with $n$ and approach the SP value of each polytype
(given in Table~\ref{table} and indicated in the plot by 
thick horizontal lines at the right). At the largest thicknesses considered,
$n$=12, the dopings amount to around 60-70\% of
their respective $P_S$ limits (see next subsection). For the
2H surface the plot shows certain curvature for $n>10$ suggesting that the
doping may saturate at a value well below $P_S$. Indeed, Fig~\ref{dosk}
shows that $\mu$ approaches the  valence band maximum (VBM) as the hexagonality
of the polytype increases due to a larger BB at the surface. For the 12~BLs
thick 2H slab, $\mu$ is already pinned at the VBM
(Fig.~\ref{dosk}(d)) and, therefore, the substrate bands
become an additional source of hole doping which competes with the graphene.

\subsection{Macroscopic model}
\begin{figure}[!hbt]
\includegraphics[scale=0.4]{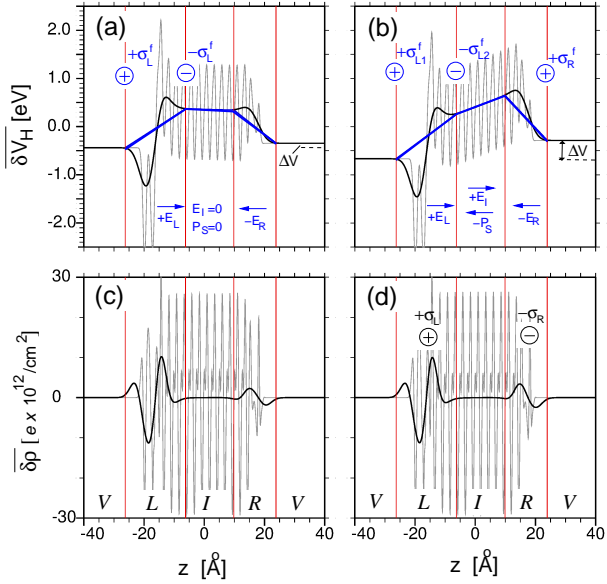}
\caption{(a) and (c) Macroscopically averaged potential and charge density
     profiles, $\vhav{z}$ and $\rhoav{z}$, respectively, for a 
     G/H/(3C-SiC)$_{12}$/Au$_2$ slab. 
     (b) and (d) same as (a) and (c) but for a 
     G/H/(2H-SiC)$_{12}$/Au$_2$ slab. 
     The macroscopic averaging has been performed via gaussian smearing 
     employing a width of 3~\AA. Thin lines in all plots show the original
     profiles averaged over the 2D unit cell but before macroscopic 
     averaging. The $V/L/I/R/V$ regions into which the system is split are
     delimited by the vertical lines, while in (a) and (b) thick blue lines
     depict the model potential profile used to analyze the data together
     with the location of the free charge sheets (only in (b)).
     In (d) total net charges in $L$ and $R$ are also sketched.
\label{macro}}
\end{figure}

We end this section by presenting a {\it macroscopic} analysis of our
results in order to rationalize the $\delta\sigma(t)$ behavior shown in
Fig.~\ref{dope} as well as to establish its connection with the
expected $P_S=-\delta \sigma$ macroscopic relationship~\cite{pdope}.
To this end, we display in Fig.~\ref{macro} with thick dark lines macroscopic 
averages~\cite{Note1} of the charge redistribution profile, $\rhoav{z}$, and 
its Hartree potential, $\vhav{z}$, for a G/H/(SiC)$_{12}$/Au$_2$ slab 
considering a 3C (a)-(c) and 2H 
(b)-(d) stacking --the original profiles are also shown as thin lines
in each plot. Both slabs present a region at the center of the dielectric where
charge neutrality is preserved ($\rhoav{z}$=0) and the local field remains
constant ($\Eav=-\partial \vhav{z}/\partial{z}$).
Thus the entire system may be
split into four sections, namely: a central (neutral) dielectric region 
({\it I}), the left and right edges ({\it L} and {\it R}) which will be 
metallic due to the G at {\it R} and the Au layers at {\it L}, and the vacuum 
region {\it V} where both $\rhoav{z}$ and the electric field vanish.
The widths of each region, $d_{I/L/R}$, are determined by requiring 
$\sigma_I$=0 and $\sigma_L=-\sigma_R$, where $\sigma_i$ is
the total charge per unit area in region $i$. The model satisfies
Gauss' law $\sigma_{R/L}=\mp \epso E_I$, with 
$E_I = \left( (\vhav{R}-\vhav{L}\right)/d_I$ being 
the local electric field in the dielectric (see Figs.~\ref{macro}(a) and (b))
generally denoted as the depolarization field in the context of
ferroelectricity~\cite{noguera,dawber}. Its origin is the incomplete
compensation of the SP by the graphene doping charge so that net charges
of opposite signs reside at each surface of the slab (in Fig.~\ref{macro}(d)
we have a net negative charge at $R$ and positive at $L$).

\begin{figure}[!htb]
\includegraphics[scale=0.45]{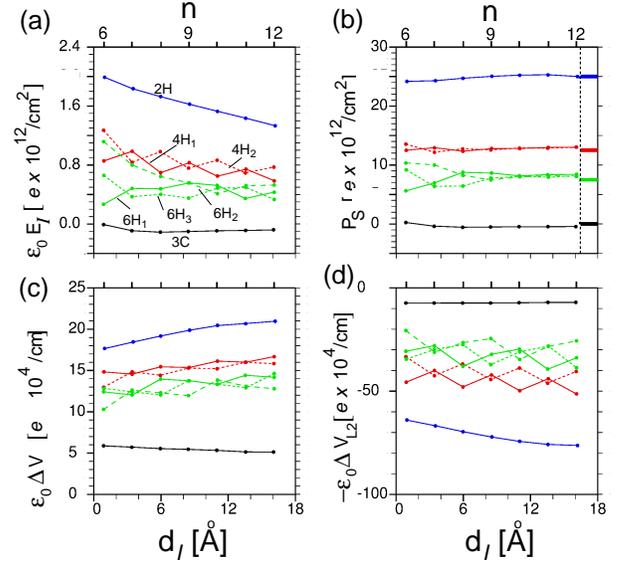}
\caption{(a) Macroscopic depolarization field $E_I$ as a function of
 the width of the dielectric region, $d_I$, for the 2H, 4H$_{1/2}$ and 
 6H$_{1/2/3}$ G/H/(SiC)$_n$/Au$_2$ slabs. 
 Color scheme same as in Fig.~\ref{1x1}.
 Solid circles correspond to the
 values derived after macroscopic averaging and lines to the fitted values
 employing eq.\eq{ei}.
 (b) SP value for each polytype and slab thickness deduced from fits to
 eq.~\eq{ei1}. (c) Calculated $\epso\Delta V$ potentials and (d) fitted
 $\epso\Delta V_{L2}$ potentials using eq.~\eq{ei} for the same cases as in
 (a) and (b). Note that, by convention, both $P_S$ and $\Delta V_{L2}$ are
 defined as positive quantities in this subsection.
 \label{fit}}
\end{figure}

Finite positive depolarization fields consistently appear for the rest of slabs
and polytypes, as shown in Fig.~\ref{fit}(a) where we plot the $E_I$ 
dependence on $d_I$. For the 2H case we obtain an
almost linear behavior indicating that the depolarization field should
vanish at large $n$ (after linear extrapolation this should occur at 
$d_I\gtrsim  50$~\AA\ or 30~BLs). The 4H slabs also show an
overall decrease as $d_I$ increases with upwards jumps when the added BL 
follows a cubic stacking while the 6H cases do not show such clear trends
specially at small thicknesses mainly due to inaccuracies in the 
determination of $E_I$.
After comparison against Fig.~\ref{dope}, a clear anticorrelation is again
found between the depolarization field $E_I$ and the doping charge 
$\delta \sigma$. Indeed, applying the continuity of the displacement field
across region $R$, both quantities are related via the SP:
\begin{equation}\label{eq:ei1}
 \delta\sigma = -\epso \epsilon_I E_I + P_S
\end{equation}
Therefore, using the data shown in Figs.~\ref{dope} and \ref{fit}(a) for 
$\delta\sigma$ and $E_I$, respectively, and assuming $\epsilon_I$ attains,
for each polytype, the bulk high frequency permittivity given in 
Table~\ref{table}, one may estimate $P_S$ as a function of $d_I$. The 
resulting SPs, shown in Fig.~\ref{fit}(b), remain essentially constant and
very close to the bulk SPs derived in the previous section (indicated
by thick horizontal lines at the right of the plot). Only the 6H slabs at 
the smallest thicknesses present substantial deviations due to the
accuracy problems mentioned above. We thus conclude that the SP has basically 
fully developed in all our model slabs so that, according to
eq.~\eq{ei1}, it is the depolarization field alone which reduces the 
graphene doping charge to values below the bulk SP.

In order to gain further insight into the origin of $E_I$, we follow
a similar approach to that employed by Dawber~{\it et al}~\cite{dawber} to 
study the effect of the electrode's thickness on the depolarization field in
ferroelectric slabs. However, at contrast with their model for the
potential profile where two thin sheets of {\it free} charge are placed at
the edges of the slab, our slab geometries are less symmetric and require
at least three such sheets, as indicated in Fig.~\ref{macro}(b) where the model
profile is superimposed by thick blue lines. One sheet is placed
at the right edge of region $R$ ($+\sigma^f_R=\delta\sigma$) 
simulating the doped graphene layer and another two 
($+\sigma^f_{L1}$ and $-\sigma^f_{L2}$), located at the left and right 
edges of region $L$, in order to model the double gold layer. 
The model grasps the main peculiarity of the slabs' macroscopic profiles 
(dark thick lines) which is the large dipole at the gold region $L$ --already
evident from the $\rhoav{z}$ profiles in Fig.~\ref{macro}(c-d). Obviously, 
the sum of these charges should yield no net free charge in the system. 
Effective constant fields $E_{L/I/R}$ may then
be defined within each region, while the potential drop in the vacuum region 
due to the DDC boundary conditions may be written as:
$\Delta V= E_L d_L + E_I d_I - E_R d_R$.
Throughout this subsection we follow the convention that all variables take 
positive values while the $\pm$ signs take care of the direction of the fields.
Employing the above relation and applying the continuity equation for the 
displacement field at the $V/L$ and $L/I$ interfaces we obtain:
\begin{equation}\label{eq:ei}
\epso E_I = \frac{P_S\;d_m + \epso( \Delta V -\Delta V_{L2})}
                {d_I + \epsilon_I\; d_m}
\end{equation}
where $d_m=d_L/\epsilon_L + d_R/\epsilon_R$ is a distance that only depends on
the width and nature of the left and right electrodes and
$\Delta V_{L2}=\sigma^f_{L2}\;d_L/\epsilon_L$ represents the potential
drop for a capacitor of charged $\pm \sigma^f_{L2}$ sheets with a dielectric
of width $d_L$ and relative permittivity $\epsilon_L$ inserted between them. 
Eq.\eq{ei} recovers the correct $E_I\approx 0$ limit as $d_I$ increases
with the slab thickness. It is essentially the same as that deduced by
Dawber~{\it et al} except for the extra $-\epso\Delta V_{L2}$ term in the
numerator. We note, however, that reasonable fits for $E_I$ using eq.\eq{ei}
cannot be achieved if this extra term is ignored.


Unfortunately, the $\Delta V_{L2}$ potential drop introduces too many unknowns 
at the $L$ electrode (namely, $\epsilon_L$ and $\sigma_{L1/2}$), which
cannot be all determined from the available computed data. Instead, we set a 
large value for the relative permittivity in this region, $\epsilon_L=10^3$ and use
eq.\eq{ei} to estimate $\Delta V_{L2}$ (the conclusions are hardly 
affected by the precise permittivity value as long as it attains reasonably
large values $\epsilon_L>20$, as expected for a metallic electrode). 
In Figs.~\ref{fit}(c) and (d) we present the
computed $\epso\Delta V$ and optimized $\epso\Delta V_{L2}$ values,
respectively, as a function of $n$ for all polytypes. 
In the absence of SP, the condition for a vanishing depolarization field
becomes $\Delta V_{L2}=\Delta V$. Indeed, in the previous subsection we
showed that by elongating the Au-Au and Au-C spacings at the bottom of the 
3C slab the graphene doping was removed (Fig.~\ref{au}e).
Within our simple model this is equivalent to increasing $\Delta V_{L2}$ at 
the cost of $\Delta V$ until both quantities equal. As shown in 
Figs.~\ref{fit}(c-d), when the same geometrical boundary conditions at the 
bottom of the slab are applied to a polytype with a finite SP, both potential 
values increase in magnitude with the hexagonality of the SiC and $n$ but, 
interestingly, $\epso\Delta V_{L2}$ attains values three to four times larger
than $\epso\Delta V$, that is, of the same order as the $P_S\;d_m$ term in
eq.\eq{ei} ($d_m\approx 5$~\AA). Therefore, and at least for the considered 
thicknesses, the depolarization field $E_I$ derives from a delicate balance 
between the left electrode's dipole contribution, $\epso\Delta V_{L2}$, and the 
sum of the SP and $\epso\Delta V$ terms.

\section{Final discussion and conclusions}
We have quantitatively studied, at the DFT level, the impact of the
bulk spontaneous polarization of the SiC substrate on the graphene's
electronic properties. First, we have presented a self-consistent scheme to
calculate the SP for polar surfaces based on standard DFT slab calculations
under DDC boundary conditions.
The scheme relies on the slab's dipole moment and its dependence on the
slab thickness, requires no macroscopic averaging and represents a 
generalization of previous works~\cite{sr,fu} as it allows to determine the
dielectric properties $\epsr$ and $P_S$ under the same eq.\eq{m5}.
The derived values are in good correspondence with previous works, while
we estimate their accuracy to be around 20-30\% which,
given the well recognized difficulties associated to such calculations,
seems satisfactory enough. A first source of error is the accuracy of the
calculated relative permittivities
(up to around 10\% after comparison with the experimental $\epsr$ values),
while a second more subtle source is related to the slab's geometry 
optimization. Here only the two upper and lower BLs of the slab were allowed
to relax while test calculations including all atoms in the relaxations
lead to $P_S$ values 10-20\% smaller. However, it is doubtful
that unconstrained relaxations provide more realistic values due to
anomalous dynamical contributions~\cite{ghosez}. In this sense, the approach
of Meyer~{\it et al}~\cite{meyer} proposing as appropiate boundary condition
for atomic relaxations a vanishing internal electric field ($\Eav=0$) by
applying a finite external field could improve the
accuracy although at the expense of longer computation times in the
self-consistent process.

Next, we examined different terminations X for G/H/SiC/X slab models which is a
crucial prerequisite to correctly account for the influence of the SP. We have 
chosen a slab terminated with a double Au capping layer which (i) pins the 
chemical potential at the bottom of the slab (instead of at the DP), hence it 
reasonably mimics a semi-infinite SiC substrate, 
(ii) presents a reduced slab dipole after expanding the Au-Au bonds and,
(iii) leads to  a vanishing doping of the graphene for the SD free 3C-SiC(111) 
substrate. Based on this slab model we have calculated the DP shifts and
graphene doping charges for 2H-, 4H- and 6H-SiC(0001) substrates as a function
of the slab thickness. Our results indeed confirm the experimentally observed 
$p$-doping in the graphene layer and reveal that it increases with the slab 
thickness and the hexagonality of the polytype, although remaining below 
the bulk SP value which, for each polytype, represents the upper limit
to the SP-derived doping (as dictated by macroscopic electrostatics). 
At the largest thickness considered of $n=12$, the graphene doping charge
reaches 60-70\% of the total SP, while a 100\% is expected at thicknesses
beyond 20~BLs; that is, far beyond the usual slab sizes considered in DFT
calculations, with the added disadvantage of requiring a hyperfine
$k$-sampling to correctly account for the graphene DOS.
Interestingly, for the most common 4H- and 6H-SiC polytypes, we find
certain dependence of the doping on the precise location of the SD closest
to the surface; for a given thickness the doping decreases by around
2~\sigmau\ the deeper it is buried due to the depolarization effect of the 
crystalline layers at the surface.

After analyzing the macroscopic averages of the charge
densities and the electrostatic potentials we ascribe this
slow convergence to the presence of a depolarization field arising from
incomplete charge compensation of the SP by the graphene doping. To 
understand the electrostatics in our slabs we find necessary to explicitly
consider the dipole moment of the gold capping double layer, yielding
a potential drop at the left electrode, $\Delta V_{L2}$, which varies
dynamically as the charge distribution across the slab changes (that is, 
with $n$ and $P_S$) and is the main responsible for the drastic reduction 
of $E_I$. Although in the current work we employed the same geometry at the
bottom of the slab for all polytypes, eq.\eq{ei} suggests that an 
alternative approach could be to tune $\Delta V_{L2}$ for each polytype
by further increasing the Au-Au spacings so as to achieve a vanishing
field within the dielectric and, hence, a 100\% compensation of the SP
by the graphene doping charge.

In summary, we have studied the relationship between the graphene doping and
the SiC substrate's SP in QFG surfaces from first principles calculations. 
Our findings suggest the possibility to tune the level of the graphene's 
doping almost in a continuous way by manipulating the number and location of 
the SDs closest to the surface. The results should naturally 
apply as well to ultrathin SiC films.


\begin{acknowledgements}
This work was supported by the Spanish Ministry of Innovation and Science
under contract Nos. MAT2013-47878-C2-R and MAT2012-38045-C04-04.
J.S. acknowledges Polish Ministry of Science and Higher Education for financing
the postdoctoral stay at the ICMM-CSIC in the frame of the program Mobility 
Plus.
\end{acknowledgements}

\appendix
\section{Spontaneous Polarizations deduced from 3D unit cells}
For the sake of completeness we present an alternative estimate of the bulk SPs based on the more traditional formalism proposed by Posternak~{\it et al} in Ref.[\onlinecite{resta}] and Qteish~{\it et al} in Ref.[\onlinecite{qteish}], which is probably the simplest one since it only requires a bulk-type (3D) calculation. Due to the imposed periodic boundary conditions the electric field generated by the internal dipoles at the SDs is compensated by a (local) depolarization field across the rest of the unit cell, $-E_{SD}$. If the internal dipole is sufficiently localized to leave a substantial region of the unit cell free of dipoles, one may obtain $-E_{SD}$ from the slope $\partial V_H/\partial z$ across this region once the electrostatic potential has been macroscopically averaged.\cite{resta} The associated spontaneous polarization is then obtained via:

\begin{equation} \label{eq:sp2}
 P_S = -\epsr \epso \frac{\partial V_H}{\partial z}
\end{equation}

\begin{figure}[!ht]
\includegraphics[scale=0.60]{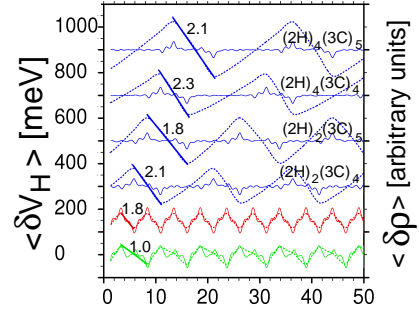}
\caption{
Macroscopically averaged charge density profiles (solid lines and right axis)
and their associated Hartree potentials (dashed lines and left axis) for
bulk calculations of different SiC polytypes. Green, red and blue lines
correspond to 6H, 4H and mixed 2H-3C SiC bulk phases, respectively. For the
latter we have considered several different supercells after varying the number
of 2H and 3C units included (indicated in each plot).
Thick straight lines overimposed on the Hartree potential ramps (depolarization
regions) correspond to linear fits whose slopes provide the  $P_S$
values indicated for each plot in \sigmauC\ units.
         \label{sp2} }
\end{figure}

As shown in Ref.~[\onlinecite{resta}], eq.\eq{sp2}\ may still be used in 
geometries where the dipoles are too close among them by constructing a larger 
supercell after adding extra SD free layers. In Fig.~\ref{sp2}(c) we plot the 
macroscopically averaged~\cite{sp1,resta} Hartree potential, $\vhav{z}$, and the
associated charge densities, $\rhoav{z}$, for the 6H-, 4H- and 2H-SiC(0001) bulk
phases.
For the latter, and since the
dipole density is large, we generated several 3D supercells 6 to 9~BLs thick 
comprising two or four 2H BLs plus four or five 3C crystalline BLs 
(see notation in the figure). 
We quote in the plots the slopes of the depolarization potentials
obtained after linear fits of $\vhav{z}$ in the crystalline regions. 
While for the 6H and 4H cases the SP values are in reasonable agreement
with those obtained from the slab calculations described in the main
text, for the 2H supercells we obtain a range of values 
$P_S^{2H}$=1.8-2.3~\sigmauC\ significantly smaller.
However, we recall certain ambiguity on the 
particular choice of $z$ at which $\partial \langle V_H\rangle /\partial z$ is 
obtained. For instance, values obtained using the local value of the partial 
derivative at the center of the ramp, or including in the fits either the 
positive or negative sections of the ramp  may lead to deviations larger than a
factor of 2 from those given in the figure.

A simple improvement of eq.\eq{sp2} is to explicitly consider the widths
of the crystalline and non-crystalline regions in the unit cell, 
$d_c$ and $d_{nc}$ respectively. The former (latter) may be 
identified with the regions where $\vhav{z}$ has a negative (positive)
slope. Applying the continuity equation for the displacement vector
across both regions we obtain:
\begin{equation} \label{eq:sp2}
 \epso (\epsilon_c E_c + \epsilon_{nc} E_{nc}) = P_S
\end{equation}
with $\epsilon_{c/nc}$ being the relative permittivities in each region and $P_S$ the SP
only present within the $nc$ section. Since the potential drop within the cell
is $\Delta \delta V_H=-E_c \;d_c = E_{nc} \; d_{nc}$, the SP may be written as:
\begin{equation} \label{eq:sp3}
 P_S = \epso \left( \frac{\epsilon_c}{d_c} + \frac{\epsilon_{nc}}{d_{nc}} \right)
                         \Delta \delta V_H
\end{equation} 
Notice that, as opposed to eq.\eq{m4}, the above expression depends on 
macroscopically averaged quantities ($\delta V_H$ and implicitly $d_{c/nc}$).
Applying eq.\eq{sp3} to the slabs shown in Fig.~\ref{sp2} and
employing the theoretically derived relative permittivities for each 
polytype given in Table~\ref{table}, we obtain:
$P_S^{6H/4H/2H}$= 1.4/2.0/2.7-2.9~\sigmauC. 
Although the agreement with the slab calculations is clearly improved
for the 2H case, the SP for the 6H and 4H polytypes is now
overestimated, suggesting that further crystalline BLs should be
included in the unit cells.

\section{Permittivity profiles}
\begin{figure}
 \includegraphics[scale=0.35]{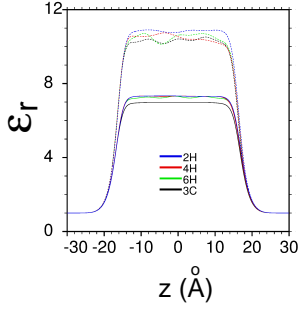}
 \caption{2D-averaged \epsrz\ profile across different
 15~BL thick SiC slabs after applying $\pm$0.1~eV/\AA\ external fields. Solid 
 and dotted lines correspond to the high frequency and static relative permittivities, 
 respectively. The profiles have been obtained after a gaussian smearing for 
 $p(z)$ employing widths of 2.5~\AA.\label{epsz} }
\end{figure}

We have additionally calculated local permittivity profiles across the dielectric
slabs, \epsrz~\cite{sr2,prb_71_144104_05}:
\begin{equation}
 \epsilon_r(z) = \frac{2 \Ee}{2 \Ee - p(z)}
\end{equation}
where $p(z)$ gives the microscopic polarization averaged
over the 2D unit cell which is obtained from the 2D averaged induced charge 
density, $\rho^{ind}(z)$~\cite{sr2}:
\begin{equation}
 p(z)= -\epsilon_0 \int_{-\infty}^z dz' \; \rho^{ind}(z')
\end{equation}
The induced charge is approximated by $\rho^{ind}(z) = \rho^+(z)-\rho^-(z)$, where
$\rho^\pm(z)$ is the charge density profile under an $\pm \Ee$ external field 
(after averaging over the 2D unit cell). In practice, large oscillations at the 
atomic scale in \epsrz\ need to be removed either by taking macroscopic 
averages~\cite{resta} for $\rho^{ind}(z)$ or employing other kind of smoothing. 
In Figure~\ref{epsz} we present \epsrz\ profiles of the high frequency 
(solid lines) and static (dashed) relative permittivities for a H/(SiC)$_{15}$/H slab, 
respectively. The profiles remain fairly constant in the inner region of the slabs
attaining values in good correspondance with those deduced above and listed in 
Table~\ref{table}. 
Notice, however, certain asymmetric features particularly in the static profiles 
probably related to anomalous dynamical contributions~\cite{noguera,ghosez}.
In fact, these features change with the slab thickness or
the precise location of the first SD in each polytype.

\end{document}